\documentclass[12pt,preprint]{emulateapj}

\slugcomment{Revised version December 2007}

\shorttitle{Spatially resolved spectra of GRB\,060505}
\shortauthors{C. C. Th\"one et al.}

\begin{document}

\title{Spatially resolved properties of the GRB\,060505 host: implications for
the nature of the progenitor \thanks{Based on ESO-ToO proposal 077.D-0661 and ESO-LP proposal 177.A-0591}
}
\author{Christina C. Th\"one\altaffilmark{1}, Johan P. U. Fynbo\altaffilmark{1}, G\"oran \"Ostlin\altaffilmark{2}, Bo Milvang-Jensen\altaffilmark{1}, Klaas Wiersema\altaffilmark{3}, Daniele Malesani\altaffilmark{1}, Desiree Della Monica Ferreira\altaffilmark{1}, Javier Gorosabel\altaffilmark{4}, D. Alexander Kann\altaffilmark{5}, Darach Watson\altaffilmark{1}, Micha\l{} J. Micha\l{}owski\altaffilmark{1}, Andrew S. Fruchter\altaffilmark{6}, Andrew J. Levan\altaffilmark{7}, Jens Hjorth\altaffilmark{1} and Jesper Sollerman\altaffilmark{1,2}}
\altaffiltext{1}{Dark Cosmology Centre, Niels Bohr Institute, University of Copenhagen, Juliane Maries Vej 30, 2100 K\o benhavn \O, Denmark}
\altaffiltext{2}{Stockholm Observatory, Department of Astronomy, Alba Nova, S-106 91, Stockholm, Sweden}
\altaffiltext{3}{Astronomical Institute 'Anton Pannekoek', University of Amsterdam, Kruislaan 403, 1098 SJ Amsterdam, the Netherlands}
\altaffiltext{4}{Instituto de Astrof\'isica de Andaluc\'ia (IAA-CSIC), P.O. Box 3.004, E-18.080 Granada, Spain}
\altaffiltext{5}{Th\"uringer Landessternwarte Tautenburg, Sternwarte 5, D--07778 Tautenburg, Germany}
\altaffiltext{6}{Space Telescope Science Institute, 3700 San Martin Drive, Baltimore, Maryland 21218, USA}
\altaffiltext{7}{Department of Physics, University of Warwick, Coventry CV4 7AL, UK}
\email{cthoene@dark-cosmology.dk}

\begin{abstract}
GRB\,060505 was the first well-observed nearby possible long-duration GRB that had no associated supernova. Here we present spatially resolved spectra of the host galaxy of GRB\,060505, an Sbc spiral, at redshift $z=0.0889$. The GRB occurred inside a star-forming region in the northern spiral arm at 6.5 kpc from the center. From the position of the emission lines, we determine a maximum rotational velocity for the galaxy of v $\sim$ 212 km s$^{-1}$ corresponding to a mass of 1.14$\times$10$^{11}$M$_\odot$ within 11 kpc from the center. By fitting single-age spectral synthesis models to the stellar continuum, we derive a very young age for the GRB site, confirmed by photometric and H$\alpha$ line measurements, of around 6 Myr which corresponds to the lifetime of a 32 M$_\odot$ star. The metallicity derived from several emission line measurements is lowest at the GRB site with 1/5 \,Z$_\odot$ but roughly solar in the rest of the galaxy. Using the 2dF galaxy redshift survey we can locate the host galaxy in its large scale ($\sim$Mpc) environment. The galaxy lies in the foreground of a filamentary overdensity extending south west from the galaxy cluster Abell 3837 at $z=0.0896$. The properties of the GRB site are similar to those found for other long-duration GRB host galaxies with high specific star formation rate and low metallicity, which is an indication that GRB\,060505 originated from a young massive star that died without making a supernova.
\end{abstract}


\section{Introduction}
GRB\,060505 reinitiated the discussion on the connection between long gamma-ray bursts (GRBs) and core-collapse supernovae (SNe) as established with the detection of a SN spectrum in the afterglow of GRB\,030329 \citep{Hjorth03, Stanek03}. Despite intense photometric and spectroscopic searches, no sign of a SN was detected for the nearby GRBs 060505 and 060614 \citep{Fynbo06, Gal-Yam06, DellaValle06}. This raised the question of whether all long GRBs are accompanied by SNe \citep{Zeh04} or whether our understanding of the explosion mechanism is incomplete \citep{Gehrels06, Fryer06, King07, Zhang07}.

The {\it Swift} satellite \citep{Gehrels} detected GRB\,060505 on May 5 2006, 06:36:01 UT, which had a fluence of (6.2$\pm$1.1$)\times$10$^{-7}$erg cm$^{-2}$ \citep{Hullinger06}. With a duration of T$_{90}$=4\,s and a statistically significant spectral lag measured from Suzaku data \citep{McBreen07} it falls in the class of long-duration GRBs. The satellite did not, however, slew automatically as the GRB was too faint to be detected in-flight \citep{Palmer06} due to a high background caused by approaching the SAA. One of the two X-ray sources inside the BAT error circle detected by the XRT was finally established to be fading \citep{Conciatore06}. An optical afterglow \citep{Ofek06} was found 1\farcs5 from the center of the revised XRT error circle with a radius of 2\farcs5 \citep{Butler06},  thereby localizing GRB\,060505 to a region 4\farcs3 north of the center of the 2dFGRS spiral galaxy TGS173Z112 at a redshift of $z=0.089$. Later imaging and spectroscopy established the burst position to be coincident with a bright, compact star forming region in one of the spiral arms of the host galaxy \citep{Thoene06, Fynbo06}, which was revealed to be a late-type, strong emission line galaxy from the 2dFGRS data \citep{Colless01}.

Host galaxies of GRBs are usually too distant to allow spatially resolved analysis with ground-based observations. Therefore, we only have information on the global properties of the galaxies, which appear to be mostly young, irregular, star forming dwarf galaxies \citep{LeFloch03, Christensen}. However, we know very little about the properties of the actual explosion sites. A recent study of GRB hosts observed with the {\it HST} \citep{Fruchter06}, which spatially resolved the host galaxies of 42 long GRBs, showed that the explosion sites coincide with the brightest regions in their host galaxies. Only two GRB hosts could so far be well resolved from the ground, those of GRB 980425 \citep{Sollerman05} and GRB\,020819 \citep{Jakobsson05}, both being spiral galaxies. In both cases, the GRBs occurred close to HII regions in the spiral arms of their host galaxies which supports the connection between long GRBs and the deaths of massive stars. 

 In this paper, we present deep observations of the host galaxy of GRB\,060505 in order to compare the galaxy and the burst site with the host galaxies of other long GRBs and to explicate the controversial nature of this SN-less long duration GRB. In \S 2, we present spatially resolved spectroscopy of the host galaxy as well as photometric data. \S 3 describes the general properties of the galaxy, concerning the classification, photometry of the entire galaxy, colors, mass determination as well as dynamical measurements. In \S 4, we examine the differences between several parts of the galaxy including the GRB site, the different stellar populations and their ages as well as differences in metallicity and extinction along the galaxy. The last section, finally, studies the large scale structure around the host galaxy.
 
Throughout the paper we adopt a cosmology with $H_0 = 71$~km~s$^{-1}$~Mpc$^{-1}$, $\Omega_{\rm m} = 0.27$, $\Omega_\Lambda = 0.73$. A redshift of z=0.0889 then corresponds to a luminosity distance of 401 Mpc and 1\arcsec{} corresponds to 1.64~kpc.

\section{Observations}
Spectra were taken with FORS2 at the Very Large Telescope (VLT) on Cerro Paranal in Chile on May 23 2006, 18 days after the burst when there was no contribution from the afterglow. We used grism 300V, which covers the wavelength range 3500 $-$ 9600 \AA{}, and a 1\farcs0 wide slit resulting in a nominal resolution of 11 \AA{} full-width-at-half-maximum (FWHM) or 590 km~s$^{-1}$ at $\lambda$ = 5600~\AA{}. Seeing conditions were decent with a seeing of 0\farcs75 FWHM, determined from the acquisition image. In order to minimize the effects of atmospheric dispersion, the FORS instrument uses a so called ``Longitudinal Atmospheric Dispersion Corrector'' (LADC) which reduces differential slit loss. We obtained two 1800 s and one 600 s exposures which were combined and reduced with standard packages in IRAF. The dispersion solution used for the wavelength calibration had an RMS of about 0.1 \AA.

\begin{figure}
\includegraphics[width=\columnwidth, angle=0]{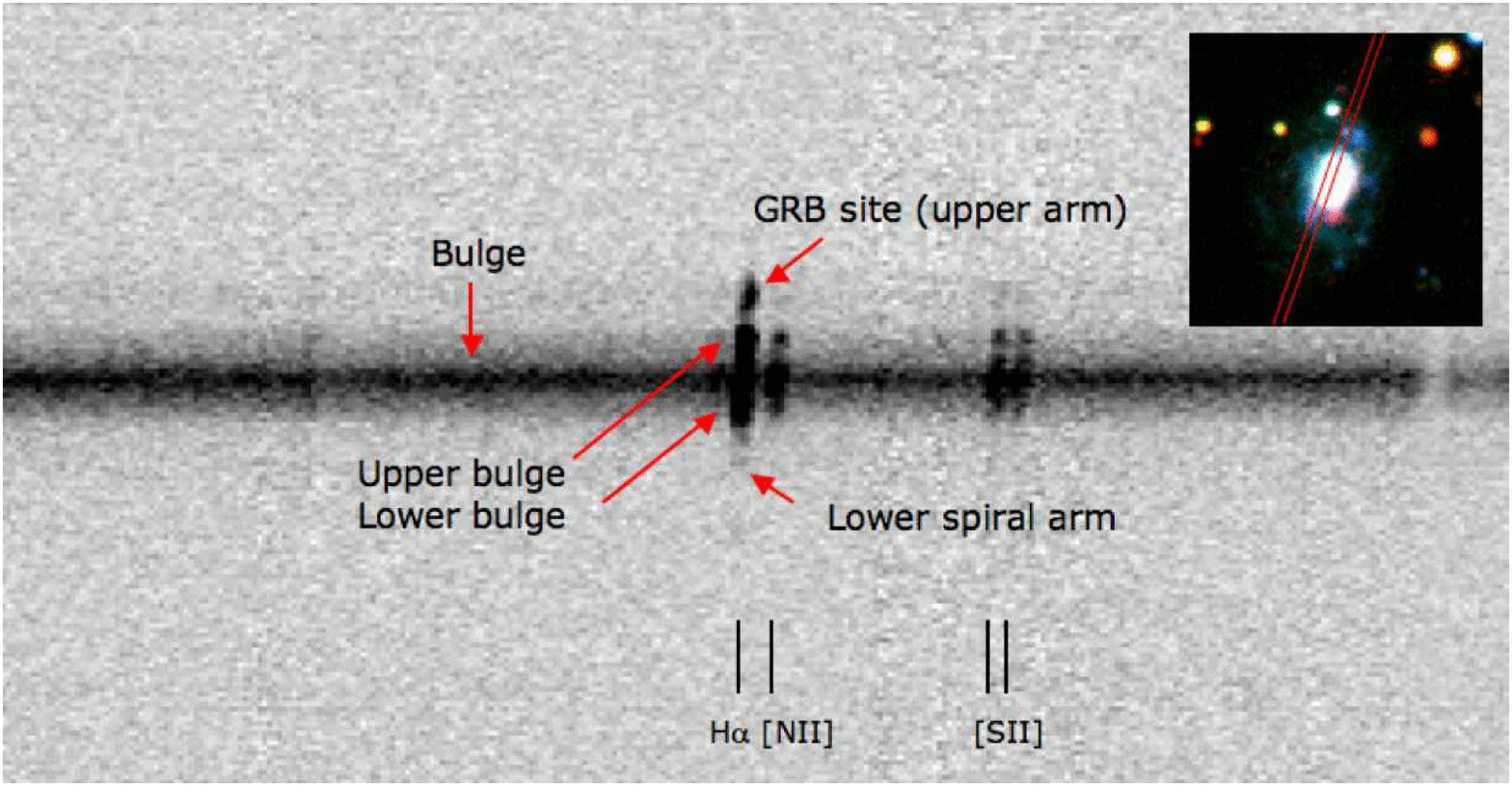}
\ \\
\includegraphics[width=\columnwidth, angle=0]{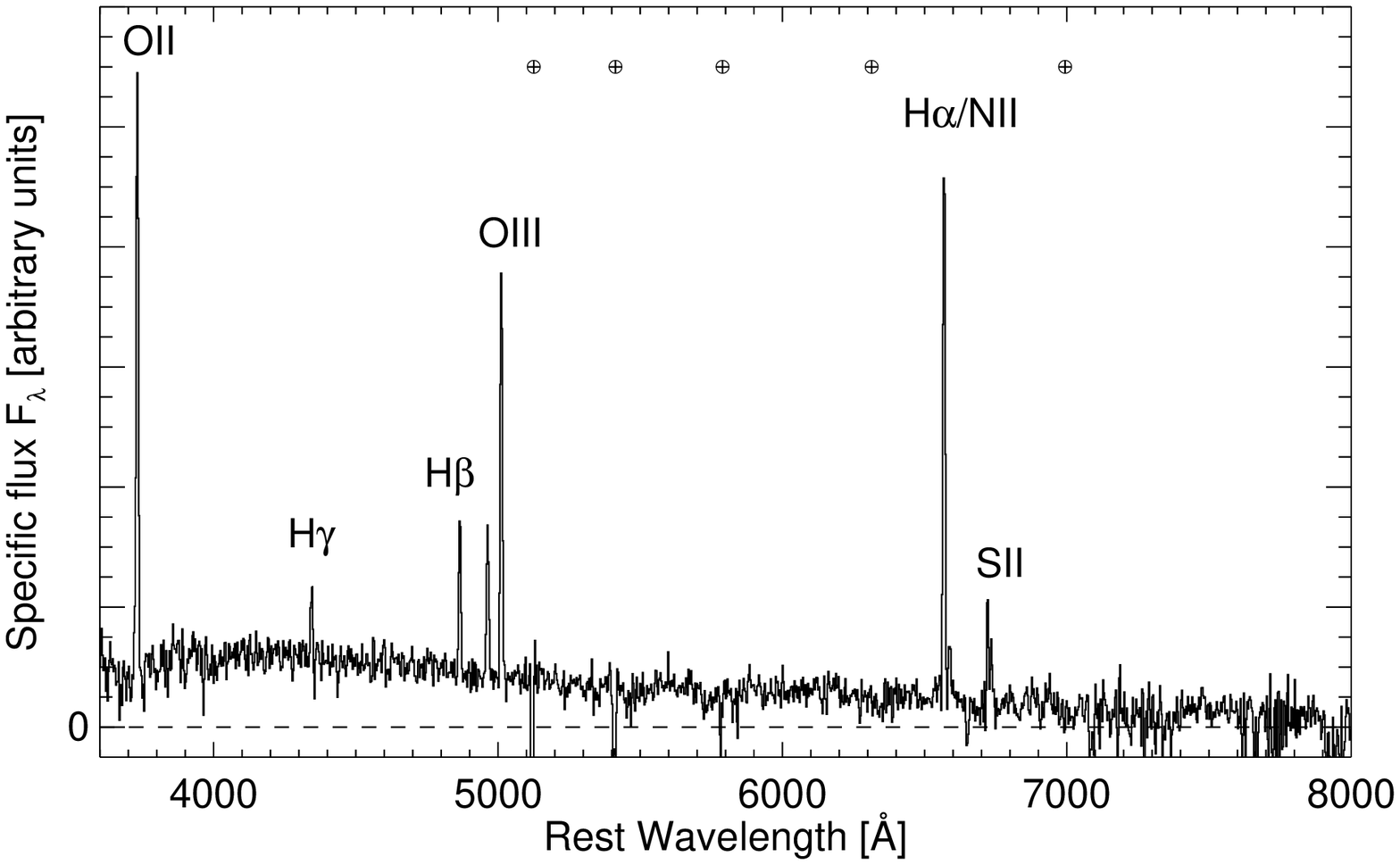}
\caption{Upper figure: 2D longslit spectrum of the host galaxy of GRB\,060505 around the H$\alpha$ emission line. Indicated are the 5 parts that were extracted from the spectrum. The inset shows the position of the slit across the host galaxy. Lower Figure: 1D spectrum at the GRB position, the crossed circles mark telluric lines. 
\label{2Dspec}}
\end{figure}

The coadded 2D spectrum was divided into four pieces of eight and one piece of nine pixels width along the spatial direction, where 1 pixel corresponds to 0\farcs25 or 0.41 kpc at the redshift of the galaxy. These five parts represent distinct regions in the galaxy such as the spiral arms and the GRB site covered by the slit (see Figures~\ref{2Dspec} and Fig.~\ref{cuts}). We used the continuum of the bulge which has the brightest trace to create a ``template'' trace and extract all five pieces using the same trace function, in order to account for the bending of the trace towards the blue. The individual parts were then flux calibrated with IRAF using observations of the spectrophotometric standard star LTT7379 from 2006 April 2 which was taken under photometric conditions. Cross-calibration with observations of the standard star LTT1788 on August 17 gave consistent results for the fluxes. We estimate the error of the flux calibration to be around 10\% in the wavelength range between 4000 and 7500 \AA{} which covers the range of the host galaxy emission lines. The flux calibration obtained in such a way can, however, only serve as a relative flux calibration and to determine the shape of the continuum.

Imaging was obtained with FORS1 at the VLT in the $BVRIz$ bands on 2006 September 14 and in $U$ on 2006 October 1 under photometric conditions. Images in the $K_S$ band were taken with ISAAC at the VLT on September 24. The images of FORS1/VLT in $UBVRI$ were calibrated using photometric zeropoints from the same night as the observations in the corresponding bands. For the $z$ band, which has no zeropoints available, we calibrated a standard field observed on the same night with magnitudes from SDSS observations of the same field, which was then used to derive instrumental zeropoints in the $z$ band. The $K_S$ band image was calibrated using a comparison star in the same field from the 2MASS catalogue \citep{Skrutskie06}.

\section{Global properties of the Galaxy}
\begin{figure}
\includegraphics[width=\columnwidth, angle=0]{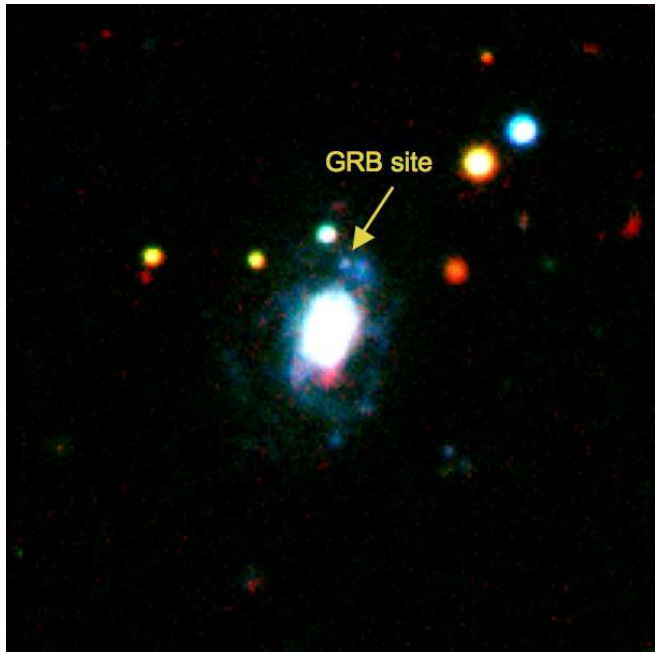}
\caption{Color picture of the host galaxy of GRB\,060505 from $BRK_S$ bands, field of view 40\arcsec $\times$ 40\arcsec, North up, East left, the position of the OT is marked.
\label{color}}
\end{figure}

\subsection{Classification}\label{Classification}
The host galaxy of GRB\,060505 is a late-type spiral galaxy with at least two major spiral arms, the GRB occurred in the northern arm of the galaxy (see Fig.~\ref{2Dspec}). From the morphology and from the strength of the H$\alpha$, H$\beta$ and the forbidden nebular emission lines (see Sect. \ref{spatial}), this galaxy can be classified as an Sbc spiral \citep{Kennicutt92}. We detect Ca H \& K absorption lines and the 4000 \AA{} break, Ca H is, however, only clearly detected in the bulge region. 
The three other spiral host galaxies of long GRBs are also late-type spiral galaxies with GRB\,980425 occurring in an SBc  spiral \citep{Fynbo00}, 990705 in an Sc \citep{LeFloch02} and GRB\,020819 presumably in an Scd spiral galaxy \citep{Jakobsson05}.

We determined the magnitudes of the entire host galaxy using the images in $UBVRIz$ from FORS/VLT and in $K_S$ from ISAAC. In order to get the total flux of the host galaxy, aperture diameters of 15\arcsec\, were used for the $U$ band, 12\arcsec\, for $BVRIz$ and 8\farcs88 for the $K$ band which were the smallest sizes to contain the total flux. The different aperture sizes were determined from a curve-of-growth analysis for each band. The flux of the three stars inside the apertures contributing about 10\% to the total flux was subtracted. These magnitudes and the corresponding colors are listed in Table \ref{hostphotometry}, the magnitudes are corrected for the foreground extinction of E(B$-$V)$=$0.021. From the $R$ band magnitude we derive an absolute magnitude for the host of M$_\mathrm{R}$~=~$-$20.15 $\pm$ 0.02~mag (no K-correction applied) which corresponds to 0.4 L$^*$ with M$_\mathrm{R}$*= $-$21.21~mag \citep{Blanton01} using h$=$ 0.71. The colors of this galaxy are actually too blue for an Sbc spiral galaxy but rather resemble the values for an irregular galaxy \citep{Fukugita95}. Also the equivalent widths (EWs) of the emission lines are generally stronger than expected for a normal Sbc spiral \citep{Kennicutt92b}. 

\begin{deluxetable}{lclr}
\tablewidth{0pt} 
\tablecaption{Photometry of the host galaxy}
\tablehead{\colhead{Filter} & \colhead{mag} & \colhead{color} & \colhead{mag}}
\startdata
U & 18.43 $\pm$ 0.05 & & \\
B & 18.89 $\pm$ 0.02 & U$-$B & $-$0.46 $\pm$ 0.05\\
V & 18.27 $\pm$ 0.02 & B$-$V & 0.62 $\pm$ 0.03\\
R & 17.90 $\pm$ 0.02 & V$-$R &  0.37 $\pm$ 0.03\\
I & 17.51 $\pm$ 0.02 & R$-$I & 0.39 $\pm$ 0.03\\
z & 17.29 $\pm$ 0.08 & R$-$z & 0.59 $\pm$ 0.08\\
K & 15.85 $\pm$ 0.04 & R$-$K & 2.05 $\pm$ 0.04\\
\enddata
\label{hostphotometry}
\tablecomments{Magnituds given are in the Vega system. The values are corrected for the foreground extinction of A$_\mathrm{V}$=0.06}
\end{deluxetable}

A closer look at the morphology reveals some asymmetry in the spiral structure and distortions in the western part of the galaxy which can also be seen in the HST images of the host galaxy presented in \cite{Ofek07} (see also Fig.~\ref{hst475}). This suggests a recent minor merger event which could have triggered the excess star formation in parts of the host which overall has an older stellar population (see Sect. \ref{pop}). It might also explain the deviation of the colors and emission line strengths compared to usual Sbc spiral galaxies.

\subsection{Measurement of the rotation curve}\label{rot}
In order to measure the rotation curve of the host galaxy, the four brightest emission lines in the 2D spectrum were used, namely [O\,{\sc ii}], H$\beta$, [O\,{\sc iii}] $\lambda$ 5008 and H$\alpha$ (see Fig.~\ref{2Dspec}). A 2D continuum subtracted postage stamp spectrum was produced for each emission line, with the continuum being modeled as a linear function fitted near each emission line. For each spatial point along the slit, the postage stamp spectrum, a Gaussian was fitted using the IRAF task {\tt ngaussfit} from the STSDAS package. At first the FWHM was kept as a free parameter, resulting in a typical value of 9$\,${\AA}. The FWHM was then fixed at 9$\,${\AA}, and a Gaussian was fitted again to each row, now with only two free parameters, the center (i.e.\ observed wavelength) and amplitude. Uncertainties on the fitted parameters were calculated using {\tt ngaussfit} based on an input noise spectrum which was calculated as photon noise and read-out noise from the 2D galaxy spectrum before sky subtraction.

The fitted observed wavelengths $\lambda_\mathrm{obs}$ were first corrected for a zero point error in the wavelength calibration and then used to calculate rest-frame line-of-sight velocities as $v_\mathrm{rest}^\mathrm{l.o.s.} = c (z - z_\mathrm{sys}) / (1 + z_\mathrm{sys})$, with $z = \lambda_\mathrm{obs}/\lambda_\mathrm{rest} - 1$ where a systemic redshift of $z_\mathrm{sys} = 0.0889$ was used. The velocities were finally corrected for inclination (deprojected velocities) as $v_\mathrm{rest}^\mathrm{deproj.} = v_\mathrm{rest}^\mathrm{l.o.s.} / \sin i$, with $i$ being the inclination (see Sect.~3.3). The row number in each postage stamp spectrum was transformed into a spatial coordinate with respect to the continuum center. The location of the continuum center as a function of observed wavelength (the ``trace'' of the continuum) was measured in a number of bins in an aperture of width 14 pixels = 3$\farcs$5 and fitted using a linear function. The used systemic redshift was chosen so that the median velocity of the 16 data points (4 per emission line) located within $\pm0\farcs5$ of the continuum center was zero, therefore, the calculated velocities were defined to be zero at the continuum center.

\begin{figure}
\includegraphics[width=\columnwidth, angle=0]{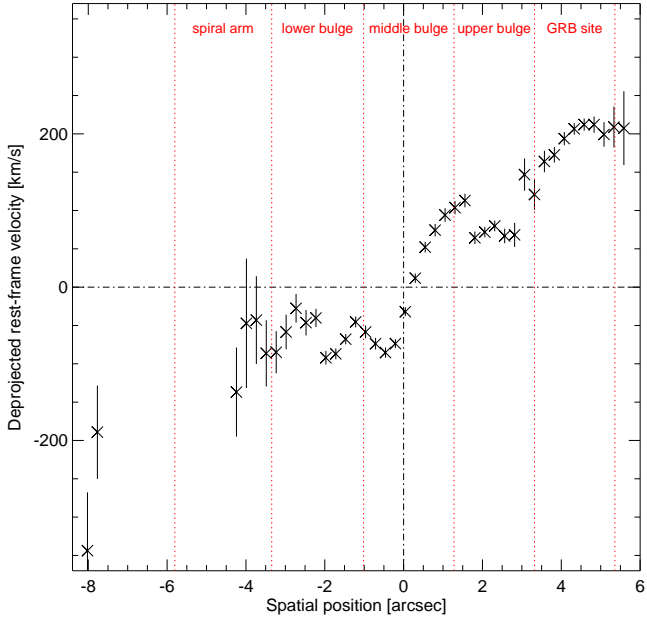}
\caption{Rotation curve using the weighted mean of the center of the emission lines [O\,{\sc ii}], H$\beta$, [O\,{\sc iii}] and H$\alpha$ over the spatially resolved host galaxy spectrum. The curve shows the true rotation curve of the galaxy, corrected for the inclination of 49 deg.
\label{rotationcurve}}
\end{figure}

To plot the rotation curve we only use points for which the fitted Gaussian amplitude was larger than three times its uncertainty. The rotation curves based on the 4 emission lines agreed reasonably well, although there were places where the difference was larger than what the calculated uncertainties could explain. The weighted mean rotation curve shown in Fig.~\ref{rotationcurve} was calculated using inverse variance weighting.

\subsection{Galaxy size and mass}
In order to determine the inclination of the galaxy, we measured the ellipticity of the disc with SExtractor \citep{Bertin96} using the photometric data in the V, R and I bands. We find e~=~0.346 $\pm$ 0.006 which gives an inclination of 49 $\pm$ 1 degrees.  The radius of the circle containing 80\% of the light is about 4\farcs 3 which corresponds to a line-of-sight radius of $\sim$11 kpc.

Furthermore, we determined the stellar and baryonic mass of the galaxy by fitting empirical models for GRB host galaxies to the spectral energy distribution (SED) of the galaxy according to \cite{Michalowski07}. This method is based on the radiative transfer code GRASIL developed by \cite{Silva98} which models the spectrum of the galaxy taking the stellar UV output and its absorption and redistribution by dust. The best fitting model was found to be similar to the SED of the GRB\,000210 host galaxy which then gives a total stellar mass of (7.9 $\pm$ 0.4)$\times$10$^9$ M$_\odot$ and a total baryonic mass of two times the stellar mass. The error quoted comes only from the errors of the model fitting, the total error from the model itself is around a factor of two. The measurement of the rotation curve which flattens at a value of 212 km~s$^{-1}$ (considering the inclination) allows an estimate of the dynamical mass to 1.14$\times$10$^{11}$ M$_\odot$ within a radius of 11 kpc.

\section{Spatially resolved properties}
\label{spatial}

\begin{figure*}
\centering
\includegraphics[scale=0.80, angle=0]{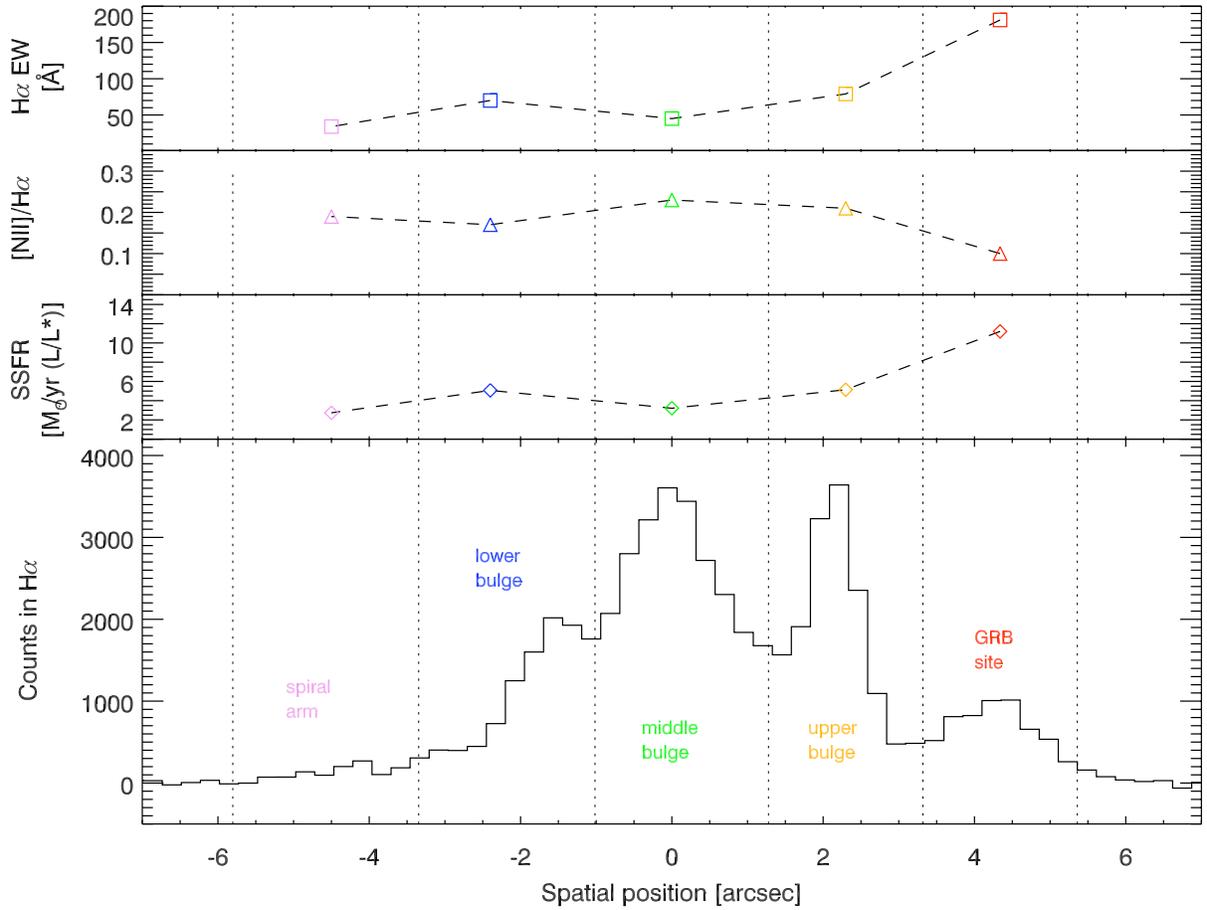}
\caption{Lower panel: Cut through the spectrum along the H$\alpha$ line and indication of the regions selected for the individual spectra (see also Fig.\ref{2Dspec}). The upper three panels show the metallicity proxy using the ratio [NII]/H$\alpha$ as described in Sec.~\ref{metallicity}, the specific star formation rate (SFR) per luminosity and the H$\alpha$ EW as a proxy for the maximum age of the youngest stellar population in the five parts of the spectrum (the higher the EW, the younger the population). All three panels show that the GRB site is considerably different from the rest of the host galaxy.
\label{cuts}}
\end{figure*}

 Fig.~\ref{cuts} shows the different regions selected for the analysis of the properties in different parts of the galaxy. The three peaks in H$\alpha$ within the bulge seem to come from the nucleus of the galaxy and the innermost regions of the two spiral arms as can be seen in Fig.~\ref{2Dspec}. The region in the upper spiral arm around the site where the GRB occurred is a large HII region, representing a clear peak in the H$\alpha$ spatial profile. A second HII region is to the west of the GRB position, just outside the slit. The spiral arm below the bulge only has a very weak peak in H$\alpha$, which indicates a low star formation rate in this region.
 
We analyze the properties of the ISM in the individual regions by comparing the emission line fluxes from the different regions (see Table \ref{lines}). The Balmer lines H$\alpha$, H$\beta$ and H$\gamma$ were found in emission in all regions, except that H$\gamma$ was not detected in the faint spectrum of the lower spiral arm. We also detect the forbidden lines [O\,{\sc ii}]\,$\lambda\lambda$3727,3729 and [O\,{\sc iii}]\,$\lambda\lambda$5007,4959 as well as [N\,{\sc ii}]\,$\lambda$6568 and the [S\,{\sc ii}]\,$\lambda\lambda$6716,6731 doublet. We measured the line properties (see Table~\ref{lines}) using the {\tt splot} task in IRAF which fits Gaussian to the lines. The lines are unresolved within the resolution of the instrument. The continuum has been corrected for the Galactic extinction of E(B$-$V)$=0.021$ mag, measured fluxes were then both corrected for the underlying stellar absorption and extinction in the host galaxy.

\begin{deluxetable}{lllll} 
\tablewidth{0pt} 
\tablecaption{Emission lines in different parts of the GRB\,060505 host galaxy}
\tablehead{\colhead{Line} & \colhead{Site} & \colhead{$\lambda_{obs}$} & \colhead{EW} &\colhead{Flux (corr.)}\\
\colhead{} & \colhead{} & \colhead{[\AA]}  & \colhead {[\AA]} & \colhead {[10$^{-17}$erg/cm$^2$/s/\AA]}}
\startdata
[O\,{\sc ii}] 3727/29&GRB	& 4064.6 	& $-$152.3	& 11.2$\pm$ 0.2 	\\
		&Bu		& 4063.2		& $-$97.9		& 12.7$\pm$ 0.3	\\
		&Bm		& 4062.9		& $-$48.91	& 16.8$\pm$ 0.4	\\
		&Bl		& 4061.8		& $-$56.38	& 13.4$\pm$ 0.4	\\
		&lS		& 4061.5		& $-$43.83	& 2.87$\pm$ 0.08	\\[2mm]
H$\gamma$  4340&GRB& 4731.6	& $-$9.82 		& 1.12$\pm$ 0.02	\\
		&Bu		& 4730.5		& $-$6.85		& 1.45$\pm$ 0.03	\\
		&Bm		& 4731.0		& $-$1.796	& 0.99$\pm$ 0.03	\\
		&Bl		& 4729.4		& $-$2.598	& 0.86$\pm$ 0.03	\\
		&lS		& \nodata		& \nodata		& \nodata			\\[2mm]
H$\beta$ 4861&GRB& 5299.9 		& $-$39.20 	& 2.98$\pm$ 0.02	\\	
		&Bu		& 5298.3		& $-$27.22	& 4.53$\pm$ 0.02	\\
		&Bm		& 5297.1		& $-$8.064	& 4.27$\pm$ 0.03	\\
		&Bl		& 5296.3		& $-$12.60	& 3.35$\pm$ 0.02	\\
		&lS		& 5297.9		& $-$8.917	& 0.65$\pm$ 0.02	\\[2mm]
[O\,{\sc iii}] 4959&GRB& 5406.4	& $-$36.31 	& 2.84$\pm$ 0.18	\\
		&Bu		& 5404.4		& $-$11.32	& 2.27$\pm$ 0.20	\\
		&Bm		& 5402.9		& $-$4.016	& 2.19$\pm$ 0.13	\\
		&Bl		& 5402.8		& $-$5.758	& 1.64$\pm$ 0.12	\\
		&lS		& 5403.4		& $-$6.004	& 0.43$\pm$ 0.03	\\[2mm]
[O\,{\sc iii}] 5007&GRB& 5458.6	& $-$92.80	& 7.20$\pm$ 0.06	\\
		&Bu		& 5456.4		& $-$40.46	& 7.84$\pm$ 0.14	\\
		&Bm		& 5456.3		& $-$9.709	& 5.17$\pm$ 0.13	\\
		&Bl		& 5455.1		& $-$17.70	& 4.85$\pm$ 0.13	\\
		&lS		& 5454.2		& $-$21.75	& 1.43$\pm$ 0.08	\\[2mm]
H$\alpha$ 6567&GRB& 7154.2	& $-$181.8 	& 9.73$\pm$ 0.03 	\\
		&Bu		&  7152.6		& $-$79.32	& 12.1$\pm$ 0.06	\\
		&Bm		&  7150.6		& $-$45.71	& 21.9$\pm$ 0.09	\\
		&Bl		&  7149.6		& $-$70.33	& 12.9$\pm$ 0.04	\\
		&lS		&  7150.1		& $-$34.70	& 1.59$\pm$ 0.02	\\[2mm]
[N\,{\sc ii}] 6586	&GRB& 7175.8 	& $-$19.44 	& 0.96$\pm$ 0.03	\\
		&Bu		& 7174.5		& $-$16.45	& 2.53$\pm$ 0.06	\\
		&Bm		& 7172.7		& $-$10.62	& 4.97$\pm$ 0.09	\\
		&Bl		& 7172.3		& $-$13.08	& 2.28$\pm$ 0.04	\\
		&lS		& 7175.4		& $-$6.142 	& 0.27$\pm$ 0.03	\\[2mm]
[S\,{\sc ii}] 6716	&GRB& 7321.1	 	& $-$50.81 	& 1.60$\pm$ 0.04	\\
		&Bu		& 7319.3		& $-$23.04	& 3.04$\pm$ 0.05	\\
		&Bm		& 7318.1		& $-$10.55	& 4.58$\pm$ 0.03	\\
		&Bl		& 7317.4		& $-$14.57	& 2.34$\pm$ 0.03	\\
		&lS		& 7316.2		& $-$29.42	& 0.77$\pm$ 0.12	\\[2mm]
[S\,{\sc ii}] 6731	&GRB& 7335.6		& $-$36.25 	& 1.09$\pm$ 0.04 	\\
		&Bu		& 7334.4		& $-$20.25	& 2.64$\pm$ 0.05	\\
		&Bm		& 7332.7		& $-$7.164	& 3.10$\pm$ 0.03	\\
		&Bl		& 7331.9		& $-$9.370	& 1.48$\pm$ 0.03	\\
		&lS		& 7334.5		& $-$36.05	& 0.91$\pm$ 0.12	
\enddata
\label{lines}
\tablecomments{Abbreviations for the different parts along the slit: ``GRB'' = GRB explosion site, ``bu'' = upper part of the bulge, ``bm'' = middle part, ``bl'' = lower part, ``ls''= lower spiral arm. Observed wavelengths are not corrected for the zero point error of around 4 \AA{} mentioned in \S \ref{rot}. The fluxes are only corrected for the Galactic extinction of E(B$-$V) $=$ 0.021 mag before measuring the fluxes, no extinction correction in the host galaxy has been applied due to its uncertainty (see Sec. \ref{extinction}) The errors in the fluxes do not include the overall error from the flux calibration which is around 10$\%$.}
\end{deluxetable}

\subsection{Burst location}\label{location}
The accurate location within the host galaxy could be determined for only a small number of bursts. For XRF\,020903 it was shown that the GRB occurred close, but not inside, a massive, star-forming supercluster \citep{Bersier06, LeFloch06, Soderberg04}. GRB\,980425 was coincident with a smaller star forming region \citep{Fynbo00, Sollerman02}. GRB\,990705 whose spiral host could be resolved in {\it HST} images also occurred close to a star-forming region within a spiral arm \citep{LeFloch02}. GRB\,020819 lies close to a brighter ``blob'', possibly a star-forming region, next to a face-on spiral galaxy, whose connection to the spiral galaxy has yet to be proven spectroscopically \citep{Jakobsson05}. 

\begin{figure}
\includegraphics[width=\columnwidth, angle=0]{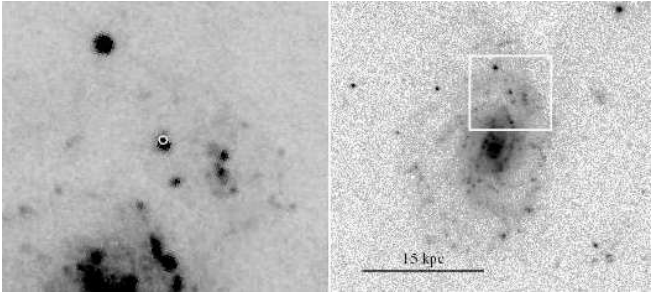}
\caption{HST image of the host galaxy of GRB\,060505 (right panel) and a smaller region showing the HII region at the position of the GRB (left panel), the optical afterglow position from ground based observations is indicated by the white circle.
\label{hst475}}
\end{figure}

{\it HST} images of the host galaxy \citep{Ofek07} show clearly that the burst lies at the edge of an HII region with a size of 400 pc. We have re-analysed these {\it HST} images, which are now public, taken in the F475W filter on May 19th, 2006 ({\it HST} program 10551, PI Kulkarni). Performing astrometry relative to the VLT observations taken on May 5th, we find that our position (see Fig. \ref{hst475}) differs slightly from that of \cite{Ofek07}, but is consistent within the errors on each measurement. Following the method of \cite{Fruchter06}, we determine that the burst occurred at the 80th percentile of the host galaxy light (i.e. 80\% of the light of the host galaxy is contained in pixels of lower surface brightness than that containing the burst) which is close to the median observed for LGRBs. The difference in the values obtained by us and \cite{Ofek07} comes from the slightly differing astrometry and the application of a small convolution to the image to allow for the positional uncertainty in our analysis. We note however that both our and Ofek's positions are consistent with the peak of the nearby HII region, which is one of the brightest regions of the galaxy. The peak of the light of this region lies at the 96th percentile of the galaxy light distribution. The absolute magnitude ($-$19.6 mag) and physical size of the galaxy ($R_{80}$) are slightly larger and brighter than the mean of LGRB hosts studied by Fruchter et al. (2006), but fit comfortably within the observed range.

The position of GRB\,060505 within the host makes it unlikely that the GRB was a chance superposition with the galaxy and in fact occurred at a higher redshift, which could have been an explanation for the lack of the detection of a SN. This clear association of the OT position with a star forming region also supports the suggestion that GRB\,060505 was due to the collapse of a massive star that originated in this star forming region. Short bursts have also been found in star forming galaxies \citep{Berger07}, but a clear association with a star forming region has not been possible so far.

\subsection{Extinction}\label{extinction}
Another issue in the discussion about the absent SN in the lightcurve of GRB\,060505 is the question of extinction along the line of sight. The Galactic extinction along the line of sight of the GRB determined from the sky maps of \cite{Schlegel} is very low with E(B$-$V)$ = 0.021$ mag. Determining the extinction in the individual parts in the host galaxy is difficult because of the incertainty in the possible Balmer absorption by an underlying older population which might especially play a role in the innermost parts of the host galaxy. For the following analysis of the properties in the different parts, we mainly used emission line ratios so that extinction correction plays a minor role.

For the GRB region, we determined the extinction from the broad-band afterglow photometry using the values of the afterglow from the series of observations in the $B,V,R$ and $I$ bands from the first night with the contribution of the galaxy subtracted (D. Xu et al. in preparation). The SED gives no indication of additional reddening along the line of sight through the galaxy and the slope is consistent, within errors, with X-ray data from \emph{Swift} XRT observations. This rules out the possibility that GRB\,060505 was obscured by dust which would have prevented the detection of a SN and also argues against the suggestion that GRB\,060505 took place at higher redshift and just happened by chance at the position of this star forming region.

\subsection{Stellar population modeling of the GRB site}\label{pop}
The age of the stellar population at the GRB site is an important key to understand the progenitor nature of this SN-less GRB. We analyze the burst region with photometry for the HII region of the GRB using an aperture of radius 0\farcs4 and a sky annulus from 0\farcs6 to 1\farcs0. The small aperture was necessary since the GRB occurred in the spiral arm of the galaxy. All the images had similar seeing except for the $I$ and $K_S$ bands which were smoothed to the resolution of the BVR images before the photometry was extracted. Aperture corrections were determined from isolated well exposed but non-saturated stars in the science images. The photometry was corrected for Galactic extinction. We did a similar analysis for the other HII region to the west of the GRB site in the same spiral arm which was outside the slit in order to investigate a possible connection between the two regions in terms of their ages. The photometry of these two HII regions is presented in Table \ref{photometry}. For the second HII region, the signal in the $K_S$ band was too low to derive any magnitude.

\begin{deluxetable}{lll}
\tablewidth{0pt} 
\tablecaption{Photometry of the GRB site and the nearby HII region west of the GRB site}
\tablehead{\colhead{Filter} & \colhead{GRB site [mag]} & \colhead{HII region [mag]}}
\startdata 
$B$ & 24.31$\pm$0.13 & 24.50$\pm$0.08\\
$V$ &  23.99$\pm$0.09 & 24.10$\pm$0.08\\
$R$ & 23.84$\pm$0.07 & 24.15$\pm$0.16\\
$I$ & 23.53$\pm$0.15 &  23.45$\pm$0.08\\
$K_S$ & 21.95$\pm$0.30 & \nodata
\enddata
\label{photometry}
\end{deluxetable}

The photometric results were then compared to predictions from the spectral evolutionary synthesis models of \cite{Zackrisson01}. These models include a realistic treatment of the nebular line and continuum emission, which has been found to be important for modelling photometric data of young stellar populations \citep{Oestlin03, Oestlin07} where the nebular component may significantly affect the broadband colors. We have redshifted the spectra from the \cite{Zackrisson01} model, integrated over the instrument throughput curve and fitted these to the observed photometry. We adopted an instantaneous burst with Salpeter IMF from 0.08 to 120 M$_\odot$ and two different metallicities for the stars and gas of Z$=$0.004 (12+log(O/H) $\sim$ 8.0 of 0.2 Z$_\odot$) and Z$=$0.008 (12+log(O/H)$\sim$8.3 or 0.4 Z$_\odot$). For the extinction we used a Galactic extinction law \citep{CCM}. One degree of freedom, namely the dust reddening can be removed from the fit as both H$\alpha$ and H$\beta$ are available from the spectrum.  In order to correct for the underlying Balmer absorption before deriving the extinction, we use the models of \cite{Gonzales97}. 

The best fits were obtained with a metallicity of Z=0.008 giving an age of 9 Myr and a reddening of E(B-V)$=$0.07 mag. It should be noted, however, that in addition to an interval around 9 Myr (7--14 Myr), the 1$\sigma$ uncertainties also allow solutions in the range of 25--100 Myr. A lower metallicity of Z$=$0.004 gives a worse fit and the only allowed interval within the 1$\sigma$ limit is 25--65 Myr. In conclusion, although favouring a young progenitor, the broadband photometry in this case does not provide very tight constraints on the stellar population age. The modeling for the HII region west of the GRB site gave an even lower age of 3$-$5 Myr applying an extinction of E(B-V)=0.25 mag. Here the extinction was left as a free parameter since no spectrum covering this region was available. 
Even though we lack a more robust measurement of the age of this region from for example spectroscopic measurements of the H$\alpha$-EW, our data suggest that these two HII regions are different from the rest of the galaxy and that their onset of star formation might indeed have been triggered at approximately the same time by a minor merging event as suggested in Sect.~\ref{Classification}.

An important indicator for the age of a population, especially when it is very young, is the predicted evolution of the H$\alpha$ EW derived from the spectra. Assuming again an Salpeter IMF from 0.08 to 120 M$_\odot$ and a metallicity of Z=0.008, the EW of $\sim$ 192 \AA{} indicates an age of 6 $\pm$ 1 Myr or less at the GRB site based on the evolution of the EW derived by \cite{Zackrisson01}. This value has to be interpreted as an upper limit due to the HI covering factor leading to a possible Lyman continuum leak which could lower the strengths of the H$\alpha$ line. The measurement of the H$\alpha$-EW is not affected by calibration errors and very reliable in terms of its physical origin as it is directly related to the percentage of very young, blue stars in the region. In order to illustrate the relative difference of the stellar population age throughout the galaxy, we plot the H$\alpha$-EW in Fig. \ref{cuts} which is inversely proportional to the age of the population as a proxy for the maximum age of the youngest stellar population.

The different methods applied for the GRB region agree in a young age for the underlying population. The age derived from the H$\alpha$-EW is most reliable from the calibration point of view as photometry of the GRB region is always affected by contamination from neighbouring regions. Both methods point to a very low age of the progenitor star of about 6 Myr which corresponds to the lifetime of a 32 M$_\odot$ star. A further determination of a young age of the GRB region could be done by directly detecting the presence of Wolf-Rayet (WR) stars \citep{Wiersema}. Unfortunately the continuum flux of our spectrum is not high enough to detect WR star lines. The young ages found for this region, however, connect well with predictions from the latest stellar evolution models for the lifetimes of massive, rapidly rotating, chemically homogeneous single star progenitors \citep{Yoon05, Woosley05}. 

\subsection{Metallicity}\label{metallicity}
In order to determine the metallicity, we used the R$_{23}$ parameter \citep[first proposed by][]{Pagel79} which is a two-valued function of the metallicity. The degeneracy between the two solutions can be broken by using the ratio [N\,{\sc ii}]/H$\alpha$ \citep{Lilly03} which gives a clear preference for the lower branch solution for the GRB site and likely a preference for the upper branch in the rest of the galaxy. In the last years, a number of recalibrations of the R$_{23}$ parameter have been done. Most of the recent metallicites of GRB host galaxies have been determined using the parametrization of \cite{KD02} for the two branches which takes into account the oxygen ionization state. Using a first guess for the metallicity and a decision for the upper or lower branch solution, the metallicity then can be determined iteratively which usually converges very fast. Recently, a new recalibration has been done by \cite{Kewley07} which corrects for the known overestimation of the metallicity from the R$_{23}$ parameter using metallicity calibrations based on measurement of the electron temperature $T_e$. In the following, we give both the \cite{KD02} values and the ones using the \cite{Kewley07} correction.

An independent estimate for the metallicity comes from the so-called $O3N2$ method using the ratio $O3N2$=log((O\,{\sc iii}/H$\beta$)/(N\,{\sc ii}/H$\alpha$)) \citep[e.g.][]{Pettini04}, which is
empirically calibrated on a sample of H\,II regions with electron temperature ($T_e$) determined metallicities. Adopting the metallicities determined through the $O3N2$ parameter, the gradient throughout the host galaxy is less pronounced compared to the estimate through the $R_{23}$ method.

A final relative metallicity estimator is the flux ratio [NII]/H$\alpha$ \citep{Veilleux87} which is proportional to the metallicity. This ratio is insensitive to underlying extinction due to the small wavelength difference between the two lines. The [NII]/H$\alpha$ ratio shows a picture consistent with the metallicity derived with the R$_{23}$ and $O3N2$ parameter. Again, we find a relatively low metallicity in the GRB region and a higher one in the other parts of the galaxy. In Fig. \ref{cuts} we therefore plot the [NII]/H$\alpha$ ratio as a proxy for the relative metallicity.

The metallicity varies largely over the different regions in the host galaxy (see Table~\ref{ISM} and Fig. \ref{cuts} for the [NII]/H$\alpha$ proxy). The star forming region at the GRB site has a relatively low value \citep[0.19 to 0.54 Z$_\odot$ for the different calibrations and taking 8.69 as value for the solar metallicity,][] {Asplund04} whereas in the rest of the galaxy, the metallicity comes close to or even above solar. Here, we also have a much older stellar population which could be responsible for the enrichment of the ISM in that part of the galaxy. From these measurements, we can infer a (O/H) metallicity gradient throughout the galaxy of $-$0.09, $-$0.09 and $-$0.04 dex kpc$^{-1}$ for the R$_{23}$ KD02, R$_{23}$ Kewley 07 and O3N2 calibrations respectively. For comparison in the local universe, for the Milky Way, M33 and M51 gradients of $-$0.07 \citep{Smartt97}, $-$0.05 \citep{Magrini07} and $-$0.02 dex kpc$^{-1}$ \citep{Bresolin04} have been derived.

\begin{deluxetable}{lllll}
\tablewidth{0pt} 
\tablecaption{Metallicities in the different parts in 12+log(O/H) and Z/Z$_\odot$}
\tablehead{\colhead{Site} & \colhead{R$_{23}$ KD02} & \colhead{R$_{23}$ K07} &\colhead{O3N2}&[NII]/H$\alpha$}
\startdata
GRB & 8.36 (0.46)& 7.96 (0.19) & 8.41 (0.54)&0.10\\
Upper Bulge & 8.84 (1.41)& 8.44 (0.56)& 8.52 (0.67)&0.21\\
Middle Bulge & 8.77 (1.20) & 8.36 (0.46) & 8.56 (0.74)&0.23\\
Lower Bulge & 8.74 (1.12) & 8.34 (0.45)& 8.52 (0.67)&0.18\\
Lower Sp. arm & 8.62 (0.85) & 8.21 (0.33) & 8.47 (0.76)&0.17
\enddata
\label{ISM}
\tablecomments{KD02 and K07 refer to \cite{KD02} and \cite{Kewley07}}
\end{deluxetable}

\subsection{Luminosity weighted star formation rate}
The instantaneous star formation rate (SFR) can be obtained from the H$\alpha$ emission flux as HI is excited by UV light mainly coming from young, blue O stars which have their peak luminosity in the UV. The conversion formula for H$\alpha$ is then SFR[M$_\odot$yr$^{-1}$] = 7.9$\times$10$^{-42}$$\times 4\pi \Phi [\mathrm{erg~cm^{2}~s^{-1}}] d_L^2$  \citep{Kennicutt98} where d$_L$ the luminosity distance and $\Phi$ the H$\alpha$ line flux.

As we only cover a part of the galaxy by the slit, we cannot make a statement about the global SFR in this spiral galaxy, but only about the relative SFR between the different parts that we extracted from the spectrum. Table~\ref{ISM} lists the SFR per 8 $\times$ 5 pixels along the slit (the area covered by the individual parts of the 2D spectrum) which corresponds to an area of 3.28 $\times$ 2.05 kpc$^2$. We then scale the SFR to the $B$-band luminosity fraction compared to the luminosity of a ``standard'' galaxy of M$_{B}$~=~$-$21 mag \citep{Christensen} to get the luminosity weighted SFR for the individual regions. For the B-band magnitudes of the parts corresponding to the individual spectra, we determine the magnitude in a rectangular aperture of 8$\times$5 pixels (9$\times$5 for the lower spiral arm) from the VLT host image at the same position as the spectrum pieces (see Table~\ref{SFR} and Fig.~\ref{cuts}). The luminosity weighted SFR is a factor of 2 to 3 higher in the star forming region around the GRB site with 11 M$_\odot$yr$^{-1}(L/L_*)^{-1}$, compared to the value in the bulge of around 3 to 5 M$_\odot$yr$^{-1}(L/L_*)^{-1}$. The latter value is higher than in an average spiral galaxy \citep{Kennicutt98} and a factor of 8 higher than in the spiral arm at the opposite side of the galaxy which has 2.7 M$_\odot$yr$^{-1}(L/L_*)^{-1}$. The luminosity weighted SFR at the GRB site, however, fits very well into the average value found in GRB host galaxies of 9.7 M$_\odot$yr$^{-1}(L/L_*)^{-1}$ \citep{Christensen}. To determine an absolute value for the SFR is, however, difficult as the absolute flux calibration might not be fully reliable. In addition, the $B$-band of our galaxy is shifted slightly towards the V-band, so the absolute magnitude in rest frame $B$-band would be smaller and therefore the luminosity weighted SFR slightly lower.

\begin{deluxetable}{lccc}
\tablewidth{0pt} 
\tablecaption{Star formation rate in the different parts}
\tablehead{\colhead{Site} &\colhead{SFR/8 pix.} &\colhead{M$_\mathrm{B}$} & \colhead{lum. weighted SFR}\\
\colhead{} & \colhead{[M$_\odot$/yr]}&\colhead{} & \colhead{[M$_\odot$yr$^{-1}(L/L_*)^{-1}$]}}
\startdata
GRB & 0.015& $-$13.82 & 11.2\\
Upper Bulge & 0.018& $-$14.86 & 5.14\\
Middle Bulge & 0.033& $-$16.03 & 3.21\\
Lower Bulge & 0.020& $-$14.99 & 5.07\\
Lower Spiral arm & 0.003& $-$13.60 & 2.74
\enddata
\label{SFR}
\tablecomments{M$_B$ is the absolute magnitude of the corresponding region in B band (observer frame).}
\end{deluxetable}

\section{The large scale structure around the GRB host galaxy}

The fact that the field of GRB\,060505 is covered by the 2dF survey \citep{Colless01} allows us to study the large scale structure around it.

The galactic environments of GRBs have so far not been studied much. At low redshifts \cite{Foley06} studied the field of the host galaxy of GRB\,980425, which was reported to be member of a group. However, based on redshift measurements of the proposed group members, \cite{Foley06} could establish that the host of GRB\,980425 is an isolated dwarf galaxy. \cite{Levan06} also proposed GRB\,030115 to be connected to a cluster around z $\sim$ 2.5 based on photometric redshifts. At redshifts $z\gtrsim$ 2 a few GRB fields have been studied using narrow band Ly$\alpha$ imaging \citep{Fynbo02, Fynbo03, Jakobsson05}. In all cases several other galaxies at the same redshift as the GRB host were identified, but it is not sure whether the galaxy densities in these fields are higher than in blank fields as no blank field studies have been carried out at similar redshifts. However, the density of Ly$\alpha$ emitters was found to be as high as in the fields around powerful radio sources that have been proposed to be forming proto-clusters, which would suggest that GRBs could reside in overdense fields at $z\gtrsim$ 2
(but note also that \citealt{Bornancini04} argue for a low galaxy density in GRB host galaxy environments).

\begin{figure*}
\centering
\includegraphics[scale=0.9, angle=0]{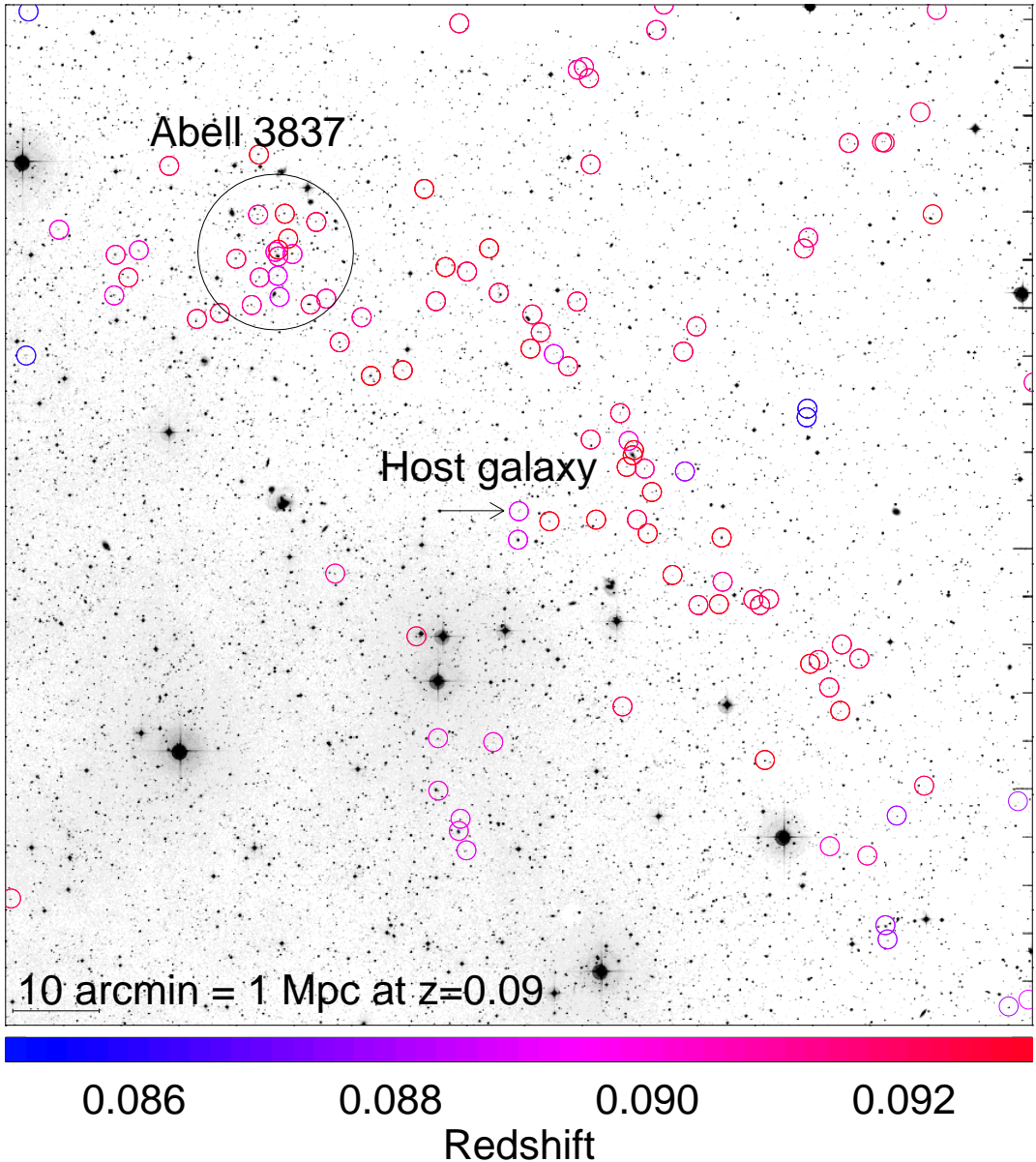}
\caption{The large scale structure around the host galaxy of GRB\,060505 shown in a field of a size of 2$^\circ\,\times\,2^\circ$. The colored circles indicate the galaxies with redshifts known from the 2dF galaxy survey. Blue indicates galaxies with a lower redshift compared to the host galaxy, red a higher redshift. Most galaxies in the field lie on a filament at slightly higher redshift than the host galaxy, stretching out from the Abell 3837 cluster to the North-East of the host. 
\label{field}}
\end{figure*}

To study the environment of the GRB\,060505 host galaxy we searched the 2dF database for all redshift measurements within a 2$^\circ \times$2$^\circ$ field around the GRB\,060505 host and with redshifts within $\Delta z = 0.004$ (about $\pm$1000 km s$^{-1}$ at the host galaxy redshift
$z_\mathrm{host}=0.089$). In Fig.~\ref{field} we show the field and the result is striking. In the GRB  field there is a large filamentary overdensity of galaxies with redshifts in the range 0.089$-$0.093. The filament extends towards the south west from the galaxy cluster Abell 3837 at $z=0.0896$ located at a
distance of 40 arcmin on the sky (4 Mpc in projection) from the host. This suggests that the host galaxy of GRB\,060505 lies in the foreground of the galaxy cluster and that it may be falling into the overdense region defined by the cluster and the filament extending out from it.

\section{Discussion and conclusions}
The low redshift of the host galaxy of GRB\,060505 puts us in the rare fortunate situation of allowing us to study the properties in the different parts of the galaxy in a spatially resolved longslit spectrum, including the star forming region around the GRB site. The spatially resolved spectrum also enables us to determine for the first time the rotation curve of a GRB host galaxy for which we find a maximum velocity of 212 km~s$^{-1}$. 

The star forming region around the burst site is shown to be different in its properties from the rest of the galaxy. Of all the parts of the galaxy traced by the slit the burst region has the highest H$\alpha$ EW, the youngest (luminosity weighted) age and the highest luminosity weighted star-formation rate and lies near the peak of the surface brightness distribution. Furthermore, the metallicity of the burst site is relatively low with only 0.19 Z$_\odot$ according to the most recent calibrations.

Connected to the nondetection of a SN in the lightcurve of GRB\,060505, there has been some discussions about the classification of the burst as a long-duration GRB \citep{Ofek07}. In addition, some concerns were put forward that the association between the GRB and the low-redshift 2dF galaxy might be due to a chance superposition \citep{Schaefer06}. The clear association of the OT position with a star-forming region with a size of 400pc, however, disfavours this possibility and strengthens the suggestion that GRB\,060505 was due to the collapse of a massive star that originated in this star-forming region. Using the size of the star forming region measured from the {\it HST} data of the host galaxy and assuming that GRB\,060505 was due to a merger event, \cite{Ofek07} derive a maximum age of 10 Myr for the progenitor system assuming the lowest possible kick-off velocity from their birth site which is consistent with the shortest time-delays of a merging system \citep{Belczynski06}. 

\cite{Ofek07} argue that the properties of the host galaxy and the location of the burst within its host provide evidence that GRB\,060505 is of a different nature than other long GRBs, in particular in comparison with the sample in \cite{Fruchter06}. As shown in this paper, the properties of the host and of the birth range fall well within the range of other long GRB hosts and birth sites in the sample of \cite{Fruchter06}. In particular, similar hosts and burst locations have been found for other nearby long GRBs. The long-duration bursts GRB\,980425 \citep{Fynbo00,Sollerman02}, GRB\,990705 \citep{LeFloch02} and GRB\,020819 \citep{Jakobsson05} were also located in the outer parts of spiral hosts. \cite{Fruchter06} also argue that even though most GRB hosts are young, metal poor dwarfs, long GRBs that are found in spiral galaxies would lie on the outskirts of those, preferring metal poor, star-forming regions which is exactly the case for GRB\,060505. 

Further support for the death of a massive star as the origin for this burst comes from the remarkably young age of the stellar population at the GRB site of less than 9 Myr, which is just barely consistent with even the shortest timescales needed for a binary compact object system to merge \citep{Belczynski06}. Also, there should be no reason to expect a low metallicity for a merging system. Even though some short-duration GRBs have occurred in star-forming galaxies \citep[e.g.][]{Fox05, Covino07, Soderberg06, Berger07} the specific star formation of the GRB\,060505 HII region is significantly larger than that inferred for these events.

\cite{Levesque07} claim that the properties of the GRB region are similar to short GRB hosts in star formation rate and metallicity derived from emission lines. Several points in their analysis, however, are highly debatable. Instead of applying a luminosity or stellar mass-weighted specific star formation rate they compare the uncorrected star formation rate with the one of entire short and long burst host galaxies. Furthermore, they take a ``sample'' of only two short GRB hosts, as their analysis requires emission lines to be able to derive metallicities for the hosts using the R$_{23}$ parameter. This leaves out a part of the short GRB host sample, namely the ones that have been claimed to reside in early type galaxies without ongoing star formation that are therefore inaccessible for metallicity measurements using R$_{23}$ or similar emission line calibrations \citep[for a summary on the properties of early type short GRB hosts see][]{Prochaska06}. In addition, one of the two bursts used in their sample is X-ray flash (XRF) 050416A, where evidence for a supernova component at late times has been found, making this burst likely to be a collapsar event \citep{Soderberg07}. Considering the still very small sample available for a detailed study of their properties and the large variety of short GRB hosts, a comparison between the hosts of the two types of GRBs should be taken with caution.

In our opinion, the evidence from the properties of the GRB region strongly suggests that GRB\,060505 was the result of the death of a massive star that died without producing a SN. Several theoretical works have developed models to explain models of GRBs that do not produce a bright or no SN at all \citep{Fryer06, Tominaga07}. Our toy-model of determining the nature of a GRB according to a (missing) SN connection seems to have to undergo a revision and be improved with additional information from the environment in order to draw conclusions concerning the GRB progenitor.

\acknowledgements
We thank the Paranal staff for performing the observations reported in this paper.
G. \"O. thanks E. Zackrisson for integrating his model spectra over the relevant filter at the observed redshift and making these models available to us. B. M.-J. wants to thank Steven Bamford for valuable discussions about rotation curve fitting of well resolved galaxies. The Dark Cosmology Centre is funded by the Danish National Research Foundation. K. W. thanks NWO for support under grant 639.043.302. D. M. acknowledges support from the Instrument Center for Danish Astrophysics. J. G. is supported by the Spanish research programs AYA2004-01515 and ESP2005-07714-C03-03.

\end{document}